\documentclass[prb,aps,twocolumn,showpacs,preprintnumbers,amsmath,amssymb]{revtex4}
\usepackage{epsf}

\begin{document}
\draft
\newcommand{\ve}[1]{\boldsymbol{#1}}

 \title{A mechanism of orbital reconstruction at the interfaces of transition metal oxides}
\author{Natalia Pavlenko}
\address{Institute for Condensed Matter Physics, 79011 Lviv, Ukraine\\
EKM, Universit\"at Augsburg, 86135 Augsburg, Germany}

\begin{abstract}
Orbital reconstruction at interfaces between YBa$_2$Cu$_3$O$_6$ and SrO-terminated SrTiO$_3$
is studied using local spin density approximation (LSDA) with intra-atomic
Coulomb repulsion (LSDA+$U$). The change of population of interfacial
Cu $3d$ orbitals results in stabilization of a new oxidation state $3d^8$
which involves an additional modification of orbital occupancies in the nearest SrO and TiO$_2$
layers. We find that an increase of electron charge in
Cu $3d_{x^2-y^2}$ states counterbalances a depopulation of $3d_{3z^2-r^2}$ orbitals
which induces, on account of the onsite Coulomb repulsion $U$, a
splitting of $3d_{3z^2-r^2}$ states at CuO$_2$-SrO interfaces.
\end{abstract}


\date{\today}

\maketitle

\section{Introduction}

Interfaces between transition metal oxides attract much attention due to their
technological 
and fundamental importance. In a complex transition metal oxide system, new electronic 
states at the interfaces between different components can bring dramatic changes 
into its physical properties. The prominent examples are metallic mixed-valence 
heterostructures of SrTiO$_3$ and LaTiO$_3$ and 
the quasi-two-dimensional electron gas at the interface between two insulators 
SrTiO$_3$ and LaAlO$_3$ with recently discussed superconducting properties \cite{ohtomo,thiel,caviglia}. 

In the transition metal oxide heterostructures, the new interfacial electronic states can be 
driven by the mechanisms of electronic reconstructions \cite{hesper}. 
These mechanisms involve charge compensation of interface polarity
or charge self-doping due to interface chemical abruptness which is one of the ways to maintain 
the overall electrostatical neutrality of the system \cite{noguera}. 
Besides the well understood 
scenarios of interface charge doping, other mechanisms of interface reconstructions can play a decisive role
in the unusual electronic properties of transition metal systems. Recently the so called orbital 
reconstruction has been shown to occur at the interfaces between cuprate and manganite films which results
in changes of $3d$ orbital occupancies of the interfacial copper ions \cite{chakhalian}. 

The first evidence for the interface orbital reconstruction 
came from resonant x-ray spectroscopy studies of 
YBa$_2$Cu$_3$O$_7$/LaCaMnO$_3$-heterostructures \cite{chakhalian}. In these experiments, the x-ray spectral intensity
at the interface between YBa$_2$Cu$_3$O$_7$ and La$_{0.67}$Ca$_{0.33}$MnO$_3$ is found to be similar for
the photon polarization in $(x,y)$-plane and in $z$ direction, which is interpreted in terms of the appearance
of hole charge in $3d_{3z^2-r^2}$ orbitals. This is in strong contrast to the standard hole doping mechanism where 
the holes occupy O $2p$ orbitals of the Zhang-Rice singlets in Cu$^{+2}$O$_2^{-4}$ planes of YBa$_2$Cu$_3$O$_{7-\delta}$.
The $3d_{3z^2-r^2}$ orbital reconstruction has been explained by a formation of a new 
``extended molecular orbital'' at the oxide interfaces in which 
a substantial amount of hole charge is distributed over the Mn, Cu and O orbitals.
As is shown in Refs.~\onlinecite{chakhalian,veenendaal}, the high Mn charge $+3$ results 
in an electron transfer from Mn to $3d_{x^2-y^2}$ orbitals of Cu, whereas the Cu $3d_{3z^2-r^2}$ states 
become partially occupied. Moreover, in such an electron doped system, the coexistence of two partially 
occupied $e_g$ orbitals leads to the formation of Zhang-Rice triplets 
centered on Cu with the total spin $S=1$. 
In contrast, for the hole-doped interfaces, the author in Ref.~\onlinecite{veenendaal} expects
a rather standard $x^2-y^2$-singlet character of the electron configuration with a stabilization
of triplet states when the hole density is increased. As is shown in Ref.~\onlinecite{veenendaal}, 
the singlet character of the electron state can be possibly changed 
due to structural Jahn-Teller distortions.
  
\begin{figure}[t]
\epsfxsize=8.0cm {\epsffile{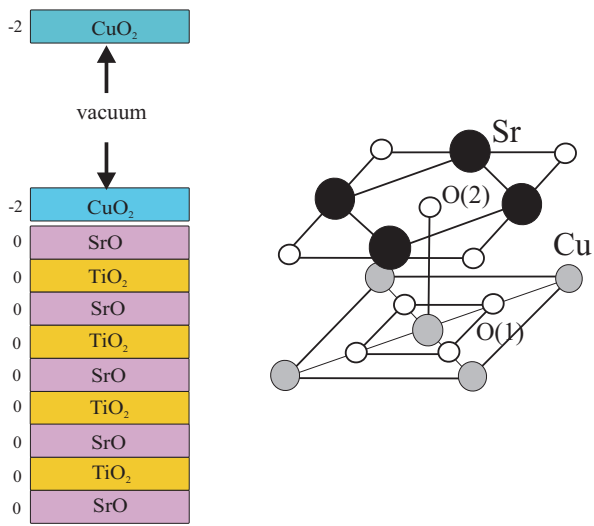}} \caption{
Scheme of a polar
CuO$_2$/SrTiO$_3$-superlattice where a STO-layer is terminated by a SrO-plane.
The right side shows a structural configuration which appears at the
interface.
} \label{fig1}
\end{figure}

In the present work, we explore theoretically a mechanism of 
the orbital reconstruction occurring through the splitting of $3d_{3z^2-r^2,\uparrow}$ and 
$3d_{3z^2-r^2,\downarrow}$ bands by the interface charge redistribution. 
As an example of a
system where the interface-caused splitting of $3d_{3z^2-r^2}$ can be obtained within the density functional
theory, we consider a supercell with a CuO$_2$ layer directly deposited on the top of 
a SrTiO$_3$ slab terminated by SrO,
a model oxide interface recently studied in Refs.~\onlinecite{pavlenko,pavlenko2} and shown schematically in Fig.~\ref{fig1}. 
In this model system, the electrostatic polarity of the interfaces between CuO$_2$ and SrO could 
be compensated via the interface self-doping by holes.
Like in the bulk YBa$_2$Cu$_3$O$_{7-\delta}$, in heterostructures
the interface self-doped holes in the CuO$_2$ layers are accumulated predominantly in the O $2p$ orbitals.
Although the charge self-doping mechanism appears to be a common scenario in a wide variety of polar 
oxide interfaces,
a close examination of the electronic structure in the CuO$_2$/SrO region unexpectedly shows
that a mechanism different from the electrostatic self-doping plays 
a key role \cite{pavlenko2}.

\section{Interface crystal fields and structural relaxation}

\begin{figure}[t]
\epsfxsize=7.5cm {\epsffile{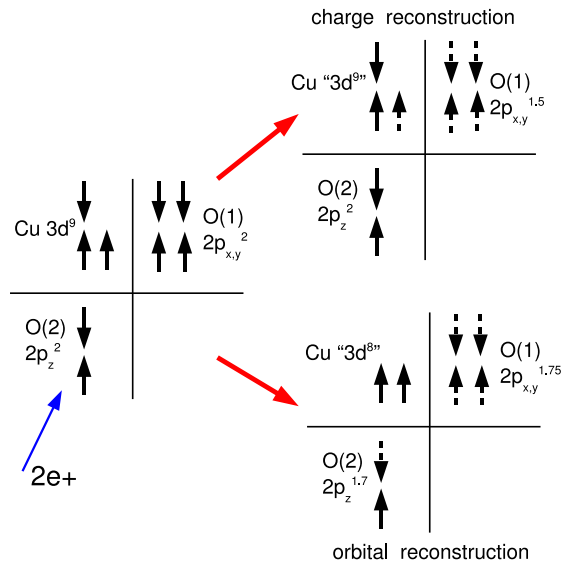}}
\caption{Schematic view of the electron configurations which 
result from hole doping.
The top and bottom right panels show the electronic and orbital reconstructed
configurations with black arrows denoting the electronically fully occupied (bold arrows)
and partially occupied (dashed arrows) orbitals. The configurations ``$3d^8$'' and ``$3d^9$''
formally describe the states $3d^8$ and $3d^9$ with partially occupied $e_g$ orbitals.}
\label{fig2}
\end{figure}

As appears from the earlier
LDA+$U$ studies \cite{pavlenko,pavlenko2}, the relaxation of the 
interfacial distance $h$ between CuO$_2$ 
and SrO planes results in the optimized value 
$h=h_0=1.83$~\AA\ which is by 0.5~\AA\
smaller than 
the distance between the apical oxygen
O(2) and Cu in CuO$_2$ planes of the bulk YBa$_2$Cu$_3$O$_6$. For larger interface distances $h\ge 2.2$\AA\,,
the compensation of the polarity proceeds through a standard self-doping mechanism of the electronic reconstruction and 
leads to an extremely high self-doping level $n_h=+2$ holes/per unit cell
which enforces charge neutrality (cf.\ the formal valency of each layer in Fig.~\ref{fig1}, to the left of the block scheme).
As can be seen in Fig.~\ref{fig2}, the holes are located in O(1) $2p_{x,y}$ orbitals (top panel, dashed arrows), 
in close analogy to the Zhang-Rice singlets described by a  $3d^9$ configuration in the CuO$_2$ planes.

\begin{figure}[t]
\epsfxsize=8.5cm {\epsffile{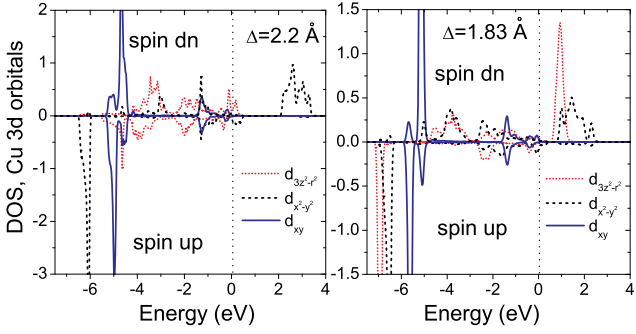}} \caption{
Partial Cu $3d$
orbital density of states at the interface between CuO$_2$ and SrTiO$_3$
terminated by SrO (LSDA+$U$ studies). The zero of energy is at the Fermi level.
The left and right panels correspond to the case of unrelaxed
($h=2.2$~\AA) and optimized ($h=1.83$~\AA) interfacial distances,
respectively.} \label{fig3}
\end{figure}

In contrast to the unrelaxed interfaces,
for the optimized distances $h~\sim h_0$ the new (orbital) mechanism of the compensation of the interface polarity becomes 
equally important. This new mechanism is associated with 
a change of the occupancy of one of 
the $3d_{3z^2-r^2}$ orbitals from one to zero due to a ``displacement'' of this band
above the Fermi level (see Fig.~\ref{fig3}). From the point of view of classical chemical valence theory,
the resulting electronic configuration can be formally described as a ``$3d^8$'' state. 
In Fig.~\ref{fig2}, the charge and orbital reconstructions caused by the $+2e$-doping level 
are schematically shown for
comparison. Formally, a
Cu $3d^8$ configuration corresponds to a change of the valency of Cu from $+2$ to $+3$ and is typically
considered as unstable in the literature. The expectation of such an instability 
was the reason for a preference for
the standard self-doping reconstruction mechanism in our earlier discussion of 
the CuO$_2$/SrTiO$_3$ system in Ref.~\onlinecite{pavlenko}.
However, a further close and careful examination of the electronic reconstruction in this system shows a 
more complex
scenario which is beyond a simple integer-like change of the electron charge on 
the copper ion (Fig.~\ref{fig2}, bottom panel). 
In this mechanism, the modification of the oxidation state of Cu can be described by two steps, namely
(i) by a depletion of electron charge from a $3d_{3z^2-r^2}$
band and from the $2p_z$ band of O(2) (SrO), and (ii) by an increase of electron density 
in the $3d_{x^2-y^2}$ orbitals hybridized with O(1) $2p_{x,y}$ states. 
These two steps provide jointly a decrease of the net electron charge on Cu which is an additional 
way to compensate the interface
polarity. Consequently, in the relaxed structure the electrostatic neutrality is achieved by a combination 
(a) of the standard self-doping mechanism with the self-doped hole charge $n_h=1$ on
O $2p$ orbitals (Fig.~\ref{fig2}, top panel), and (b) of the $3d_{3z^2-r^2}$ orbital 
reconstruction which leads to a decrease of the local 
electron charge on Cu (Fig.~\ref{fig2}, bottom panel). 

Let us estimate the crystal field which would be sufficient for stabilization
of the new oxidation state formally denoted here as extended Cu $3d^8$.
Using the notations $\{ \uparrow, \downarrow \}$ for the majority and minority spin states, the $3d^8$ state
can be described as $3d^{5\uparrow, 3\downarrow}$ configuration where the occupied states
$3d_{x^2-y^2,\uparrow}$ and $3d_{3z^2-r^2,\uparrow}$  have a character of 
a Zhang-Rice triplet, similar to the results of Ref.~\onlinecite{veenendaal}.
In the relaxed structure, this
configuration corresponds to the minimum of the total energy. In the LDA+$U$ calculations, the energy gain
$\Delta E_{tot}=E_{tot}(h_1)-E_{tot}(h_0)$ due to the relaxation of the interface distance
from $h=h_1=2.2\AA$ to $h=h_0$ approaches $1$~eV.
It is remarkable that in the extended Cu $3d^8$ configuration, the main contribution 
to $\Delta E_{tot}$ is provided by the reconstruction 
of the electronic states in the nearest CuO$_2$ and SrO planes
\begin{eqnarray} \label{dEtot}
\Delta E_{tot}=&& \Delta \varepsilon_{3z^2-r^2}+\Delta \varepsilon_{x^2-y^2}\nonumber \\
&& +\Delta \varepsilon_{xz+yz}+\Delta p_O^2,
\end{eqnarray} 
where $\Delta \varepsilon_{3z^2-r^2}$, $\Delta \varepsilon_{x^2-y^2}$,
and $\Delta \varepsilon_{xz+yz}$ 
are the changes of the $3d$ orbital 
energies; $\Delta p_O^2$ refers to the shifts of the energies of the $2p$ bands
of oxygens O(2). In our calculations, the band energies are represented by their 
gravity centers \cite{watson,eyert}. 
Here a gravity center $E_g$ of a band with a density
of states $\rho(E)$ is calculated from the expression $E_g={\int dE E \rho(E)}/{\int dE \rho(E)}$.
As the displacements of the gravity centers of 
the $2p$ states of O(1), $2p_x$ and $2p_y$ states of O(2) 
and $3d_{xy}$ bands of Cu are negligibly small as compared to the splitting of $e_g$, $3d_{xz}$ 
and $3d_{yz}$ bands, they are not included into (\ref{dEtot}). 

In the SIC LDA+$U$ approach \cite{wien2k,anisimov}, the $3d_{3z^2-r^2}$ orbital energies in the optimized ($\nu=0$) and
unrelaxed ($\nu=1$) system can be represented as
\begin{eqnarray} \label{ez2}
\varepsilon^{\nu}_{3z^2-r^2,\sigma}=
\varepsilon^{\nu(0)}_{3z^2-r^2}+U_{eff}\left(\frac{1}{2}-n^{\nu}_{3z^2-r^2,\sigma}\right).
\end{eqnarray}
Here $U_{eff}=U-J$ and $\sigma$ is the spin index. In our calculations, we consider the local Coulomb corrections
for Cu $3d$ orbitals $U=8$~eV, and the Hund coupling $J=0.8$~eV. 
In the unrelaxed structure, the states $d_{3z^2-r^2}$ are filled by 
electrons: 
$n^1_{3z^2-r^2,\sigma} \approx 1$. In contrast, in the optimized system, the orbital reconstruction leads to 
$n^0_{3z^2-r^2,\uparrow} \approx 1$
and $n^0_{3z^2-r^2,\downarrow} \approx 0$.
The quantities $\varepsilon^{\nu(0)}_{3z^2-r^2}$ are the corresponding LDA orbital potentials which are assumed to be 
independent of the spin direction. As a consequence, the total $3d_{3z^2-r^2}$ energy shift 
for the relaxed structure is given by 
\begin{eqnarray} \label{dez2}
&& \Delta \varepsilon_{3z^2-r^2} =\Delta \varepsilon_{3z^2-r^2,\uparrow}n_{3z^2-r^2,\uparrow}+
\varepsilon^1_{3z^2-r^2,\downarrow}n^1_{3z^2-r^2,\downarrow}\nonumber \\
&& \approx \Delta \varepsilon^{10}_{3z^2-r^2}+n^1_{3z^2-r^2,\downarrow}
\varepsilon^{1(0)}_{3z^2-r^2,\downarrow}
-\frac{1}{2} U_{eff}, 
\end{eqnarray}
where the potential difference $\Delta 
\varepsilon^{10}_{3z^2-r^2}=\varepsilon_{3z^2-r^2}^{1(0)}-\varepsilon_{3z^2-r^2}^{0(0)}$ 
is determined by the 
change of the orbital crystal field due to the interfacial relaxation.

From (\ref{ez2}), the change of the splitting 
of the spin bands $\varepsilon_{3z^2-r^2,\uparrow}$ and $\varepsilon_{3z^2-r^2,\downarrow}$
due to structural relaxation can be described as  
\begin{eqnarray} \label{dez_updn}
&& \!\!\! \!\!\! \!\!\!
\Delta \varepsilon_{3z^2-r^2,\uparrow}- \Delta \varepsilon_{3z^2-r^2,\downarrow}\nonumber \\ 
&&  = U_{eff} (n_{3z^2-r^2,\downarrow}^1-n_{3z^2-r^2,\downarrow}^0)
\approx U_{eff}.
\end{eqnarray}
where $n_{3z^2-r^2,\downarrow}^\nu$ is the occupation of the ($3z^2-r^2,\, \downarrow$) orbital in the optimized ($\nu=0$) and
unrelaxed ($\nu=1$) system.
The last result demonstrates that the 
on-site
Coulomb repulsion controls the spin band splitting and the consequent enhancement of the local magnetic moments. 

As the width of the $3d_{3z^2-r^2}$ bands exceeds the shift $\Delta \varepsilon_{3z^2-r^2}$
caused by the orbital reconstruction, the use of the centers of gravity
to estimate the changes of the crystal fields cannot be sufficiently accurate
to convey the physical picture. Therefore, we consider a different
analysis of these fields. In this approach, we calculate the centers of gravity of 
the $3d_{x^2-y^2}$ and $t_{2g}$  bands which are expected to be accurate due to 
their substantial shifts by
about $2$~eV caused by the crystal field. The obtained values for $\Delta \varepsilon_{x^2-y^2}$
and $\Delta \varepsilon_{xz+yz}$
together with the similarly calculated 
$\Delta p_O^2$ can be directly
substituted into the equation (\ref{dEtot}). 
For the known $\Delta E_{tot}$ this
equation provides an estimate for the minimal energy $\Delta \varepsilon_{3z^2-r^2}$
\begin{eqnarray} \label{de10z2}
\Delta \varepsilon^{10}_{3z^2-r^2}\approx  \Delta E_{tot}+\frac{1}{2}U_{eff}-\Delta \varepsilon_{x^2-y^2}\nonumber \\
-\varepsilon^{1(0)}_{3z^2-r^2}n^1_{3z^2-r^2,\downarrow}-\Delta \varepsilon_{xz+yz}-\Delta p_O^2,
\end{eqnarray}
which is sufficient for the stabilization of the orbitally reconstructed
extended Cu $3d^8$ configuration. Here $\varepsilon^{1(0)}_{3z^2-r^2}=-1.86$~eV is the LDA-energy level 
of the unrelaxed $3d_{3z^2-r^2}$ orbital.

\begin{table}
\caption{\label{tab1}Band centers of gravity (in eV) for different values of interfacial distance $h$.
Here, $\varepsilon_{2,z}$ refers to the $2p_z$ orbital of O(2).}
\begin{ruledtabular}
\begin{tabular}{llllllll}
$h$~(\AA), $\sigma$ & $\varepsilon_{x^2-y^2}$ & $\varepsilon_{xz+yz}$ &
$\varepsilon_{2,z}$ \\
\hline
1.83 ($\uparrow$) & -5.03 & -5.23 & -2.98 \\
1.83 ($\downarrow$) & -1.07 & -4.84 & -2.62 \\
2.20 ($\uparrow$) & -4.68 & -3.60 & -1.93 \\
2.20 ($\downarrow$) & 0.97 & -3.16 & -1.82
\end{tabular}
\end{ruledtabular}
\end{table}

\begin{figure}[ht]
\epsfxsize=7.0cm {\epsffile{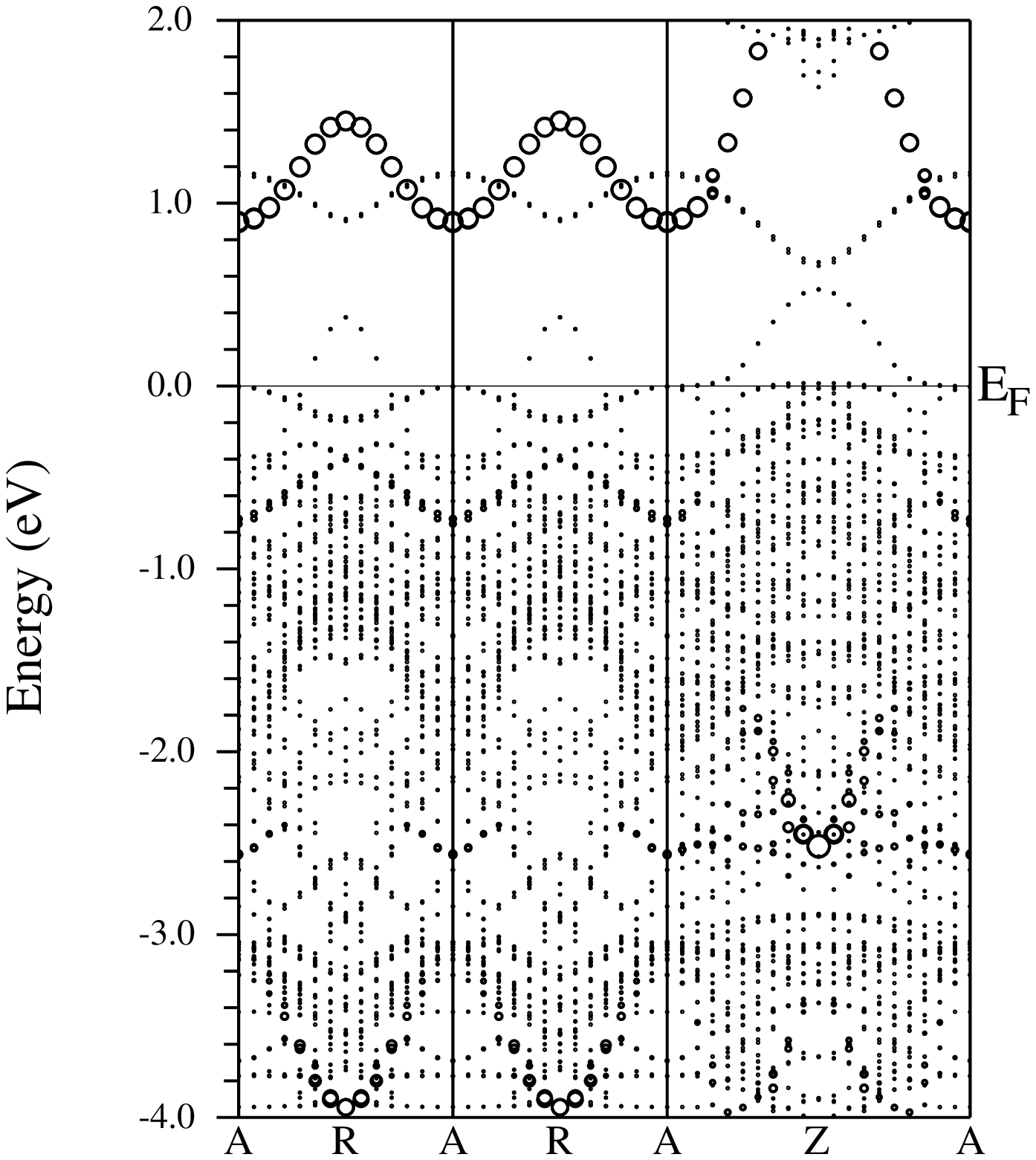}}
\epsfxsize=7.0cm {\epsffile{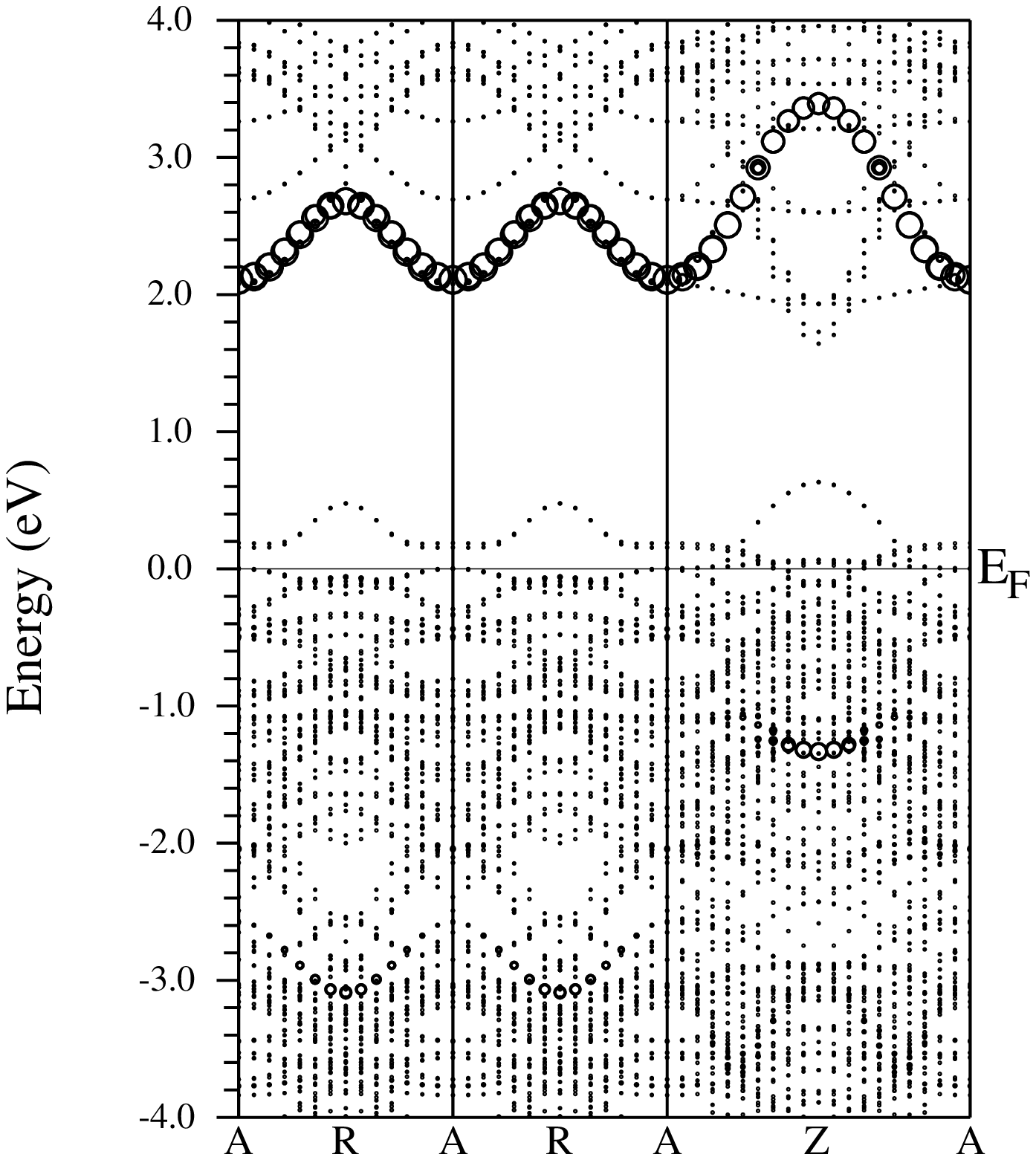}} \caption{
Band structure for different $h$ where the top panel
corresponds to $h=1.83$~\AA~and 
the bottom panel to $h=2.2$~\AA.
The size of the circles is proportional
to the Cu $3d_{x^2-y^2,\downarrow}$ character of the respective states. Here the coordinates of
the symmetric points of the Brillouin zone
are: A=($-\pi$, $\pi$),
R=($-\pi$, $0$), and Z=($0$,$0$).} \label{fig4}
\end{figure}

In expression (\ref{de10z2}), the energies of the $3d_{x^2-y^2}$ bands can be 
represented as
\begin{eqnarray} \label{ex2y2}
\varepsilon^\nu_{x^2-y^2,\sigma}=\varepsilon^{\nu(0)}_{x^2-y^2,\sigma}+U_{eff}\left(\frac{1}{2}-
n^{\nu}_{x^2-y^2,\sigma}\right).
\end{eqnarray}
In distinction to the reconstructed $3d_{3z^2-r^2}$ orbitals, the occupation numbers $n^{\nu}_{x^2-y^2,\sigma}$ 
of $3d_{x^2-y^2}$ states are affected only slightly by the decrease of $h$. In (\ref{ex2y2}), 
the latter property leads to small changes by 0.07~eV in the second local Coulomb contribution
due to atomic relaxation.
Consequently, the effect of the interface relaxation on $3d_{x^2-y^2}$ is reduced to a modification
of the LDA potentials $\varepsilon^{\nu(0)}_{x^2-y^2,\sigma}$:
\begin{eqnarray} \label{de_x2y2}
\Delta \varepsilon_{x^2-y^2}=\Delta \varepsilon^{(10)}_{x^2-y^2,\uparrow}+\Delta \varepsilon^{(10)}_{x^2-y^2,\downarrow},
\end{eqnarray}
where $\Delta
\varepsilon^{(10)}_{x^2-y^2,\sigma}=\varepsilon^{1(0)}_{x^2-y^2,\sigma}n^1_{x^2-y^2,\sigma}
-\varepsilon^{0(0)}_{x^2-y^2,\sigma}n^0_{x^2-y^2,\sigma}$.
Using the results of the LDA+$U$ calculations, the shift $\Delta 
\varepsilon^{(10)}_{x^2-y^2,\sigma}$ and the concentrations $n^{\nu}_{x^2-y^2,\sigma}$ can be found
from the centers of gravity of $3d_{x^2-y^2}$ bands and by integration of the densities of states 
obtained for the relaxed and optimized interfaces.
Similarly to $\Delta \varepsilon_{x^2-y^2}$, the shift $\Delta 
\varepsilon_{xz+yz}$ can be easily determined from the changes of the LDA potentials
$\varepsilon^{\nu(0)}_{xz+yz,\sigma}$. 

\begin{figure}[ht]
\epsfxsize=6.5cm {\epsffile{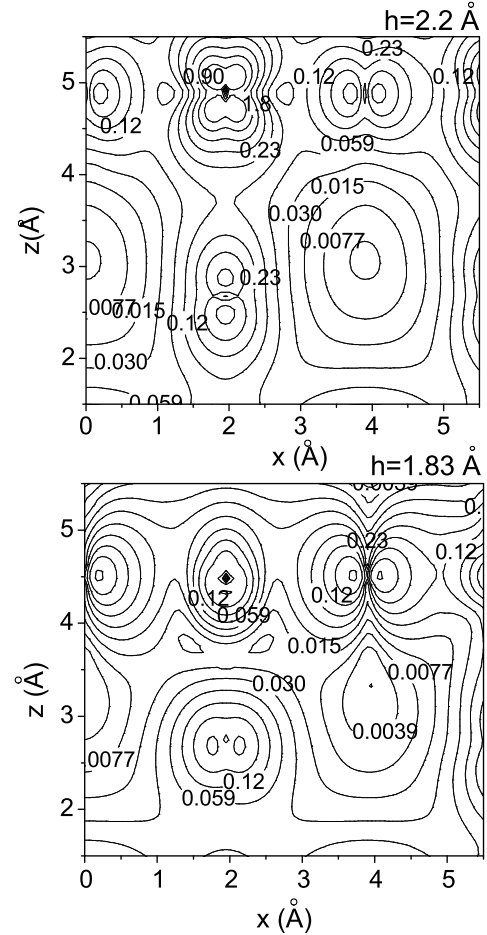}}
\caption{
Electron density map (in e/\AA$^3$) in the energy range $-4.5$~eV~$\le E$~$\le -2.5$~eV below the Fermi level
for unrelaxed and optimized interface distances. Here the electron density is plotted
in the ($x$, $z$) plane with the $x$-axis along the O(1)-Cu-O(1) diagonal of 
the CuO$_4$ plaquettes shown in the right panel of Fig.~\ref{fig1}.}
\label{fig5}
\end{figure}

The band gravity centers calculated
for Cu $3d_{x^2-y^2}$, $3d_{xz+yz}$ and for $2p_z$ orbitals of O(2) in superlattices with unrelaxed and optimized $h$ are 
presented in 
Table~\ref{tab1}. From this table, one can detect that a central feature of the
interface relaxation is a strong shift of the center of the minority band $3d_{x^2-y^2,\downarrow}$ down
to the energy $-1.07$~eV below the Fermi level. The reason
for this shift can be understood from the analysis of the band structures of
the optimized and unrelaxed interfaces plotted in Fig.~\ref{fig4}. In Fig.~\ref{fig4}, the size of the circles
is proportional to the Cu $3d_{x^2-y^2,\downarrow}$ character of the respective states.
The comparison shows a substantial Cu $3d_{x^2-y^2,\downarrow}$ contribution to the $2p$ states of 
oxygens in 
the CuO$_2$ planes which are located below the Fermi level. This contribution is especially significant
at the energies about 
$-(0.6-0.8)$~eV, at $-2.5$~eV as well as at about $-(3.8-4)$~eV. 
Specifically, the integration of the partial densities of states below the Fermi level shows 
an increase of the population $\Delta n_{x^2-y^2,\downarrow}$ of Cu $3d_{x^2-y^2,\downarrow}$ 
levels by about 0.2 electrons in the relaxed structure. 
To check the numerical accuracy of this result we have also calculated the increase of the electron concentration
from equation (\ref{ex2y2}) as 
$\Delta n_{x^2-y^2,\downarrow}=n_{x^2-y^2,\downarrow}^{0}-n_{x^2-y^2,\downarrow}^{1}=
(\varepsilon^{1}_{x^2-y^2,\downarrow}-\varepsilon^{0}_{x^2-y^2,\downarrow}
-\Delta\varepsilon^{10(0)}_{x^2-y^2,\downarrow})/U_{eff}=0.25$ with $\Delta\varepsilon^{10(0)}_{x^2-y^2,\downarrow}=
\varepsilon^{1(0)}_{x^2-y^2,\downarrow}-\varepsilon^{0(0)}_{x^2-y^2,\downarrow}$,  
which agrees well
with the value obtained by the direct integration. Besides the $3d_{x^2-y^2,\downarrow}$ state, the relaxation 
also results in an increase of the population of $3d_{x^2-y^2,\uparrow}$ orbital by about 0.1 electrons, 
as compared to 
the system with $h=2.2$~\AA.

\begin{figure}[t]
\epsfxsize=8.0cm {\epsffile{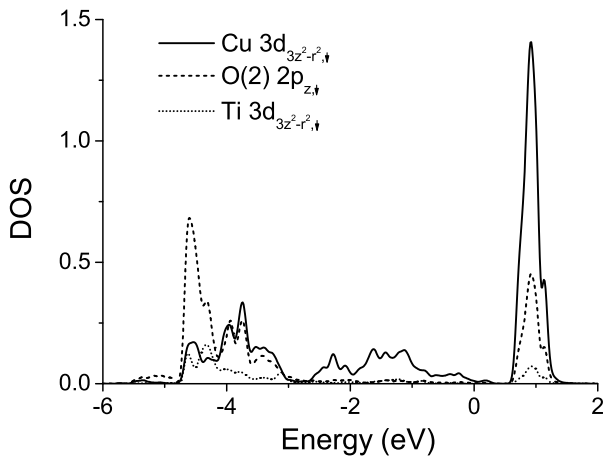}}
\caption{
Projected Cu $3d_{3z^2-r^2,\downarrow}$, O(2) $2p_{z,\downarrow}$ and Ti $3d_{3z^2-r^2,\downarrow}$
orbital densities of states at the interface between CuO$_2$ and SrTiO$_3$
in the relaxed heterostructure (LSDA+$U$ studies). The zero of energy is at the Fermi level.}
\label{fig6}
\end{figure}

Using the values of the band centers given in
Table~\ref{tab1}, we find from Eq.~ (\ref{de10z2}) that the minimal
energy shift $\Delta \varepsilon^{10}_{3z^2-r^2}$ sufficient for orbital
reconstruction is given by $\Delta \varepsilon^{10}_{3z^2-r^2} \approx 1$~eV.
As the typical values of the crystal fields expected in this type of systems can approach 
$2-3$~eV, we can conclude
that the mechanism of the orbital reconstruction realized
through the stabilization of the extended Cu 
3$d^8$ configuration shown in Fig.~\ref{fig2} (bottom panel) 
appears to be a plausible scenario. We note that this reconstruction occurs
despite a strong local Coulomb repulsion $U=8$~eV which is 
overcome by the modifications of the structure
of the Cu $3d_{x^2-y^2,\downarrow}$  and $3d_{3z^2-r^2,\downarrow}$ bands.

\section{Electronic density at orbital reconstruction}

To understand better the nature of the reconstruction of Cu $3d_{x^2-y^2,\downarrow}$, we also analyze
the ($x$, $z$) electron density maps (represented in
Fig.~\ref{fig5}) in the energy range of the bonding states Cu $3d_{x^2-y^2}$-O $2p$.
Here the $z$-axis is directed perpendicular to (001).
The decrease of the interface distance
$h$ leads to stronger electron repulsion in the nearest CuO$_2$ and SrO layers. The effect of this repulsion
is clearly visible in the bottom panel of Fig.~\ref{fig5} ($h=1.83$~\AA) which shows electron depletion
in the spatial region between
Cu (located at $x=2$~\AA, $z=4.5$~\AA) and 
oxygen O(2) ($z=2.5$~\AA) in 
the SrO-plane. The decrease of the electron density
along the $z$-direction has an extended character and involves Cu $3d_{3z^2-r^2,\downarrow}$, O(2) $2p_{z,\downarrow}$ 
and Ti $3d_{3z^2-r^2,\downarrow}$ orbitals
which may be interpreted as reduced hybridization between these orbitals. 
In Fig.~\ref{fig6} this decrease is 
manifested through a peak centered at $E=1$~eV above the Fermi level
which appears due to the splitting of 
the $3d_{3z^2-r^2}$ bands. The peak remains significant even in the density 
of Ti $3d_{3z^2-r^2,\downarrow}$ states which are more distant from the interface. The consequent increase 
of the difference between the electron occupancy of the majority and minority spin states also leads
to the enhancement of the magnetic moment of Cu from 0.65~$\mu_B$ to 0.86~$\mu_B$, 
an effect discussed in the previous section.

\begin{figure}[t]
\epsfxsize=6.5cm {\epsffile{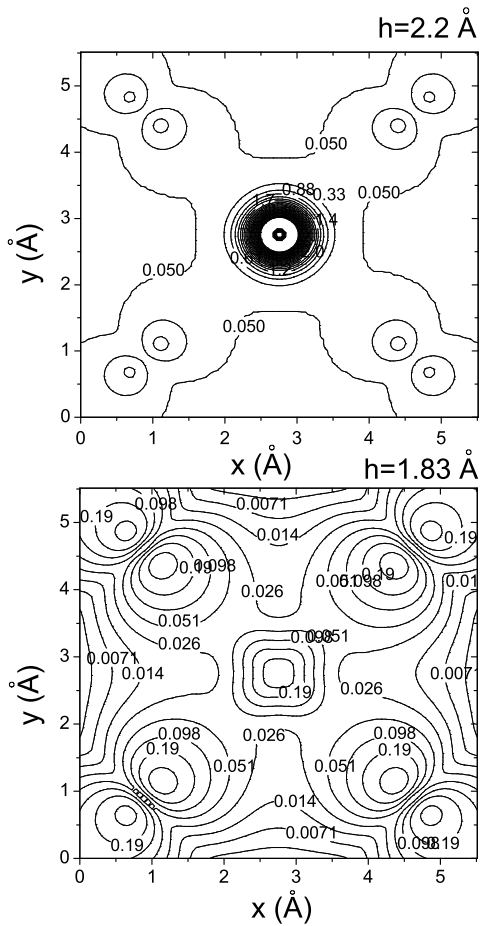}}
\caption{
Electron density map (in e/\AA$^3$) in the energy range $-4.5$~eV~$\le E$~$\le -2.5$~eV
below the Fermi level in ($x$, $y$) plane
for unrelaxed and optimized interface distances. Here $z/c=0.7$.}
\label{fig7}
\end{figure}

In contrast to the electron depletion in $z$ direction, in the CuO$_2$-plane the 
interface relaxation results in a strong charge
hybridization between Cu and O(1) which is presented in more 
detail in Fig.~\ref{fig7}. To estimate the modification of electron charge resulting from the combination 
of (i) electron depletion across the interface and (ii) electron accumulation in CuO$_2$ planes, we have calculated
planar ($x$, $y$) and orthogonal ($z$) contributions to the electron density. To obtain the orthogonal 
part (i) (electron depletion), we have 
extracted the charge density in the range between 0.5~eV and 1.4~eV which corresponds to the 
new DOS-peak. The direct integration of the obtained spatial density profiles 
results in the value $n_e^z\approx 1.45$ 
carriers per unit cell. This value should be corrected by 
$\Delta n_e^{z}$ on account of an overlap
with the states $3d_{x^2-y^2}$, $2p_x$ and $2p_y$ which are shifted towards 
lower energies in the relaxed system 
(see Fig.~\ref{fig3}). The integration of the corresponding projected DOS 
parts gives $\Delta n_e^{z}=0.28$.
It should be noted that the latter integration has been performed without 
consideration of extra charge in the interstitial regions
which should give additional contributions to $\Delta n_e^{z}$. In our analysis, the interstitial contributions
are expected to be small and the resulting $\Delta n_e^{z}$ does not exceed the maximal value 
of about 0.45--0.5.
As a consequence, the obtained modification of the depleted charge density
$\Delta n_{itf}=n_e^z-\Delta n_e^{z} \approx 1$.
This implies that approximately one
electron per unit cell is depleted from the $3d_{3z^2-r^2}$ and $2p_z$ orbitals of the interfacial bonding complex 
Cu-O(2)-Ti.
In the interfacial stack CuO$_2$-SrO-TiO$_2$, about 70$\%$ of the depleted charge is located in the CuO$_2$ layers 
which suggests a strong spatial charge confinement to the $3d_{3z^2-r^2}$ orbitals of Cu.  

To calculate the amount of the electrons accumulated in the CuO$_2$ planes of the relaxed structure (contribution (ii)), we 
performed the integration of 
the hole charge from the Fermi level to the top of 
the O(1) valence band which gives 
$n_h^{xy}=1$ holes.
These holes are located in the $2p_{x,y}$ states of O(1) hybridized with the $3d_{x^2-y^2}$ orbitals of Cu. The 
comparison with the value $n_h^{xy}(h_1)=2$ for the unrelaxed system 
provides the resultant amount $n_e^{xy}=1$ for the 
additional electron charge accumulated due to structural relaxation. 
It is remarkable that in their combination, both contributions
(i) and (ii) give the total positive charge $+2$e required for the overall neutrality of the relaxed system.

Therefore, in 
this mechanism for the orbital reconstruction
one can distinguish two key factors 
which are responsible for the stabilization of the extended Cu 3$d^8$ configuration near structurally relaxed 
CuO$_2$/SrTiO$_3$ interfaces:
(i) the depletion of the electrons from a $3d_{3z^2-r^2}$ orbital of Cu hybridized with 
$2p$ O(2) and $3d_{3z^2-r^2}$ states 
of Ti, a process caused by the splitting of the $3d_{3z^2-r^2}$ bands;
(ii) the increase of electron charge density in the Cu $3d_{x^2-y^2,\downarrow}$ orbitals which are
strongly hybridized with the $2p_{x,y}$ bands of the oxygens O(1) in the CuO$_2$ planes. This increase is
accompanied by 
a shift of the $3d_{x^2-y^2,\downarrow}$ band towards lower energies and it provides the energy gain 
sufficient for the orbital reconstruction. 
It is worth noting that the second stage 
(increase of electron concentration on O(1) $2p$ and Cu $3d_{x^2-y^2,\downarrow}$) in 
fact counterbalances the electron depopulation in stage (i).
In this way, it minimizes the lost of energy 
due to the splitting of Cu $3d_{3z^2-r^2}$ bands 
with the orbital reconstruction.
The obtained charge self-regulation is in agreement with the negative feedback mechanism discussed 
in Ref.~\onlinecite{raebiger} which supports its general character for the transition metal oxides.

\section{Conclusions}

To summarize, we note that in the orbital reconstruction mechanism,
the effective change of the valency of Cu due to the splitting of $3d_{3z^2-r^2}$ bands appears to be 
a plausible and realistic property. For the cuprate-titanate interfaces, the resulting
electronic state is characterized by the electronic configuration $3d^{5\uparrow, 3\downarrow}$
of Cu $3d$ orbitals extended by the changes of orbital occupancies of ions
in the neighbouring SrO and TiO$_2$ planes. Surprisingly, the strong hybridization
between O $2p_{x,y}$ and Cu $3d_{x^2-y^2,\downarrow}$ states of CuO$_2$ planes plays an essential role,
making even moderate crystal fields 
acting on Cu $3d_{3z^2-r^2}$ orbitals sufficient for the
orbital reconstruction at CuO$_2$-SrO interfaces and the consequential splitting of  the $3d_{3z^2-r^2}$ states 
through the onsite Coulomb interaction $U$.
An increase of the magnetic moment of Cu (from  0.65~$\mu_B$ to 0.86~$\mu_B$) is predicted for the reconstructed 
interface state.

\section*{ACKNOWLEDGEMENTS}
This work was partially supported through the DFG~SFB-484. The author thanks
Thilo Kopp for helpful discussions. Grants of computer time from the 
Leibniz-Rechenzentrum M\"unchen are gratefully acknowledged.

\end{document}